\documentclass[fleqn,twoside]{article}
\usepackage{espcrc2}
\usepackage{graphicx}

\catcode`\@=11
\def\ps@fnal{\def\@oddhead{\hfil \textsf{FERMILAB--Conf--04/109--T}}
\def\@evenhead{\thepage \hfil FERMILAB--Conf--04/109--T}}

\newcommand{\slj}[3]{\mbox{$^{#1}${\ifcase#2\or S\or 
         P\or D\or F\or G\fi}$_{#3}$}}

\newcommand{\ev}{\hbox{ eV}}
\newcommand{\kev}{\hbox{ keV}}
\newcommand{\mev}{\hbox{ MeV}}
\newcommand{\gev}{\hbox{ GeV}}

\newcommand{\pb}{\hbox{ pb}}
\newcommand{\fb}{\hbox{ fb}}

\newcommand{\cfrac}[2]{\textstyle{\frac{#1}{#2}}}
\newcommand{\jpsi}{\ensuremath{J\!/\!\psi}}

\def\ltap{\,\raisebox{-.4ex}{\rlap{$\sim$}} \raisebox{.4ex}{$<$}\,}
\def\gtap{\,\raisebox{-.4ex}{\rlap{$\sim$}} \raisebox{.4ex}{$>$}\,}

\newcommand{\AmS}{{\protect\the\textfont2
  A\kern-.1667em\lower.5ex\hbox{M}\kern-.125emS}}
\title{The Lost Tribes of Charmonium}
\author{Chris Quigg\thanks{Fermilab is operated by
Universities Research Association Inc.\ under Contract No.\
DE-AC02-76CH03000 with the U.S.\ Department of Energy.}       \address{Fermi National Accelerator Laboratory, 
  P.O. Box 500, Batavia, Illinois 60510 USA}}

\begin{document}
\begin{abstract}
To illustrate the campaign to extend our knowledge of the charmonium 
spectrum, I focus on a puzzling new state, $X(3872) \to \pi^+\pi^-\jpsi$.  Studying
the influence of open-charm channels on charmonium properties leads us
to propose a new charmonium spectroscopy: additional discrete
charmonium levels that can be discovered as narrow resonances of
charmed and anticharmed mesons.  I call attention to open issues for theory and
experiment.
\vspace{1pc}
\end{abstract}

\maketitle
\thispagestyle{fnal}
\addtocounter{footnote}{-1}%
\section{MISSING LEVELS}
New experimental results---including the discoveries of new 
states---have revitalized the study of heavy 
quarkonium~\cite{Skwarnicki:2003wn,Stoeck,Athar} and stimulated a fresh wave of 
theoretical analysis.
In this talk, I want to focus on 
three groups of elusive narrow charmonium states. 
(1) All interpretations of the charmonium spectrum anticipate 
two additional $c\bar{c}$ states below $D\bar{D}$ threshold: the 
1\slj{1}{2}{1}  $J^{PC} = 1^{+-}$ level, $h_{c}$, near the 1\slj{3}{2}{J} 
centroid, and the 2\slj{1}{1}{0}  $J^{PC} = 0^{-+}$ level, 
$\eta_{c}^{\prime}$, the hyperfine partner of $\psi^{\prime}(3686)$. 
(2) We have long expected that two unnatural parity states---the 
1\slj{1}{3}{2}  $J^{PC} = 2^{-+}$ level, $\eta_{c2}$, and the 
1\slj{3}{3}{2} $J^{PC} = 2^{--}$ level, $\psi_{2}$---would lie between 
the $D\bar{D}$ and $D\bar{D^{*}}$ thresholds. Forbidden by parity 
invariance to decay into two pseudoscalars, these states should be 
narrow in the traditional charmonium sense. (3) New coupled-channel calculations indicate 
that several levels with open charm-anticharm decay channels should 
be observable as narrow structures.

\begin{figure}[tbh] 
\includegraphics[width=2.75truein]{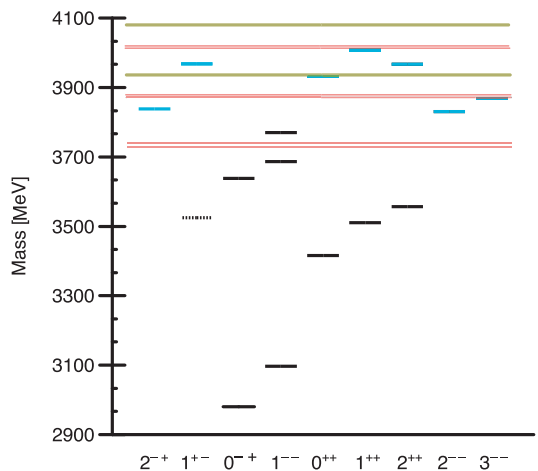}
\vspace{-24pt}
\caption{Grotrian diagram for the charmonium spectrum.  States marked
by heavy black lines are well established.  The \slj{1}{2}{1} $h_c$
level is indicated by the dashed line at the \slj{3}{2}{J} centroid.
Thresholds are shown, in order of increasing mass, for $D^0\bar{D}^0$,
$D^+D^-$, $D^0\bar{D}^{*0}$, $D^+\bar{D}^{*-}$, $D_s\bar{D}_s$,
$D^{*0}\bar{D}^{*0}$, $D^{*+}\bar{D}^{*-}$, and $D_s\bar{D}_s^*$.  Some
predicted states above threshold are depicted as faint lines.}
\label{fig:Grotrian}
\vspace*{-24pt}
\end{figure}
There is still much to be learned from the study of $c\bar{c}$ states.
Including the interthreshold
region between $2M(D)$ and $M(D)+M(D^*)$, we expect about ten or eleven
narrow levels, of which at least seven are already known.  Including
higher states within $800\mev$ of charm threshold, we expect perhaps
sixty states, to be observed either as discrete levels or through their
collective effect on the total cross section for $e^+e^- \to \hbox{
hadrons}$.  A portion of the charmonium spectrum is shown in
Figure~\ref{fig:Grotrian}.  
Nonrelativistic potential models
historically have given a good account of the spectrum, but they cannot
be the whole story.  They are truncated, single-channel treatments that
do not contain the full richness of quantum chromodynamics.  We are
coming closer to a complete theoretical treatment: lattice QCD is
increasingly capable for quarkonium spectroscopy---and improvements are
coming swiftly~\cite{Kronfeld:2003sd,Ishikawa}.  On the experimental 
side, charmonium states are being seen in electron-positron
annihilations, in $B$ decay, in two-photon collisions, and in hadronic
production.  This circumstance gives us access to a very broad variety
of quantum numbers $J^{PC}$, and makes for a lively conversation among
experiments and a fruitful dialogue between theory and experiment.

\subsection{Indications for $h_{c}$}
Twelve years ago, Fermilab experiment E760 reported evidence for
resonant formation of the 1\slj{1}{2}{1} state of charmonium in
proton-antiproton annihilations~\cite{Armstrong:1992ae}.  They saw a
narrow resonance at $3526.2\mev$ in the isospin-violating
$\pi^{0}\jpsi$ channel.  The absence of fresh news has left the $h_{c} $
in limbo.

At this meeting, Claudia Patrignani presented interesting new
results from the successor experiment, E835~\cite{Claudia}. In their 
 new data set, they find no evidence for $h_{c} \to 
\pi^{0}\jpsi$, and infer an upper limit on the product of branching 
fractions, $\mathcal{B}(h_{c}\to \bar{p}p)\mathcal{B}(h_{c}\to 
\pi^{0}\jpsi)$, that is about one-third of the E760 level. However, 
in the three-photon channel, a preliminary analysis finds 13 $\eta_{c}\gamma$ 
candidates close to the 1\slj{3}{2}{J} centroid, with an expected 
background of 1 or 2. Interpreted as 
examples of $h_{c} \to \eta_{c}\gamma$, these events would imply a
resonance mass of $3525.8\mev$, with a reasonable value of 
$\Gamma(h_{c} \to \bar{p}p)\mathcal{B}(h_{c} \to \eta_{c}\gamma)$.

This new E835 work---in the canonical $\gamma\eta_{c}$ mode---should
give added stimulus to the search for the \slj{1}{2}{1} state in other
experiments.  The cascade decay $B \rightarrow h_c K^{(*)} \rightarrow
\gamma \eta_c K^{(*)}$ offers one promising
approach~\cite{Suzuki:2002sq,Gu:2002nh,Eichten:2002qv}.
In hadron colliders, it may be possible to observe $\eta_{c} \to 
\varphi\varphi$ ($\approx 3$ per mille branching fraction) or 
another hadronic mode, with or without a secondary vertex tag to 
enhance $B$-decay as a source, then to look for the 500-MeV photon 
from $h_{c} \to \gamma\eta_{c}$.\footnote{We estimate  $\Gamma(h_{c} \to 
\gamma \eta_{c}) \approx 460\kev$ in the Cornell coupled-channel 
model~\cite{Eichten:2004uh}, 
which suggests that $\mathcal{B}(h_{c} \to \gamma\eta_{c}) \approx 
\cfrac{2}{5}$.} 

\subsection{Discovery of $\eta_{c}^{\prime}$}
Twenty years  passed without a confirmation of the Crystal Ball claim of the 
2\slj{1}{1}{0} $\eta_{c}^{\prime}(3594\pm5)$ 
\cite{Edwards:1982mq}, and the complementary technique of charmonium 
formation in $p\bar{p}$ annihilations did not support the 
$\eta_{c}^{\prime}(3594)$ evidence \cite{Ambrogiani:2001wg}. 
In 2002 came Belle's observation 
of $\eta_{c}^{\prime}$, at a higher mass, in exclusive $B \to K K_{S} 
K^{\mp}\pi^{\pm}$
decays~\cite{Choi:2002na}.  CLEO~\cite{Asner:2003wv},
BaBar~\cite{Wagner:2003qb}, and Belle~\cite{Abe:2003ja} have confirmed
and refined the discovery of $\eta_{c}^{\prime}$ in $\gamma\gamma$
collisions, fixing its mass and width as $M(\eta_{c}^{\prime}) = 3637.7
\pm 4.4 \mev$ and $\Gamma(\eta_{c}^{\prime}) = 19 \pm
10\mev$~\cite{Skwarnicki:2003wn}. It is worth noting that the 2004 
Review of Particle Physics~\cite{PDBook2004} regards the 
$\eta_{c}^{\prime}$ as needing confirmation. Let us hope for 
definitive experimental results soon!

The outstanding issue for the \slj{1}{1}{0} $\eta_c^\prime(3638)$ is the
small splitting from its \slj{3}{1}{1} hyperfine partner $\psi^\prime$,
compared to potential-model expectations, which we shall examine
presently.

\subsection{Discovery of $X(3872)$}
\begin{figure*}[tbhp] 
    \includegraphics[height=3.25truein, width=18pc]{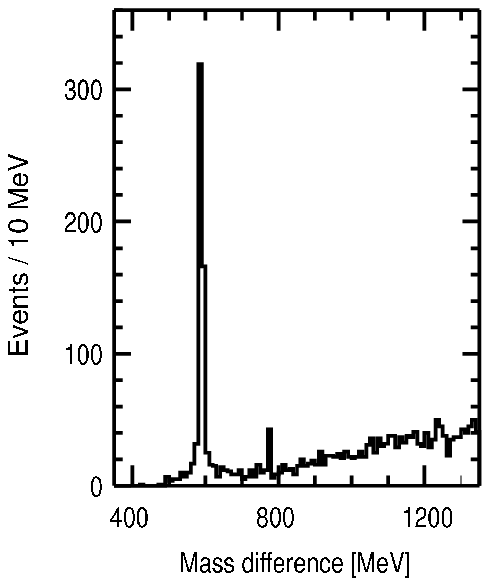}\quad 
    \includegraphics[height=3.25truein]{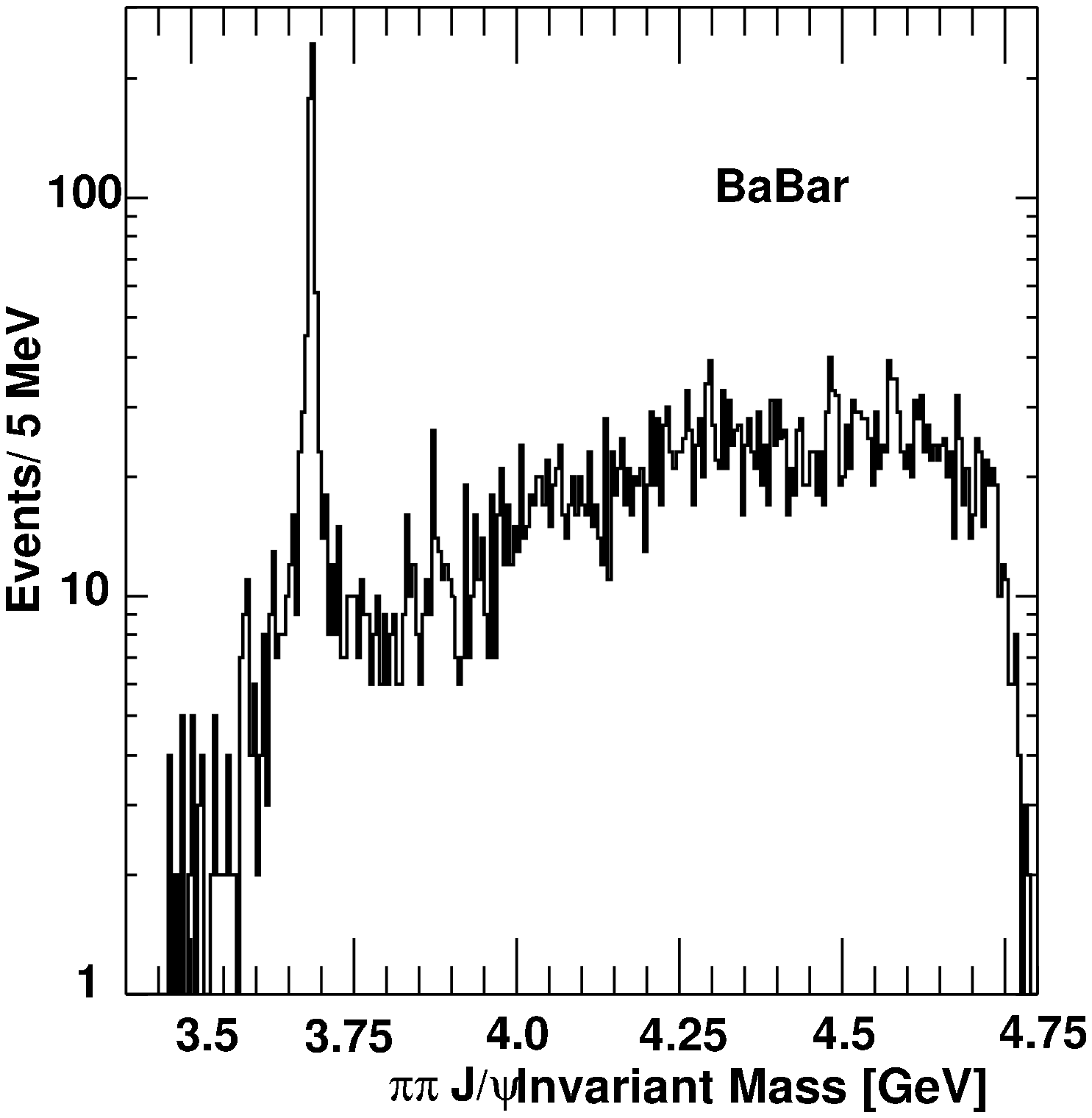}\\
    \includegraphics[height=3.5truein]{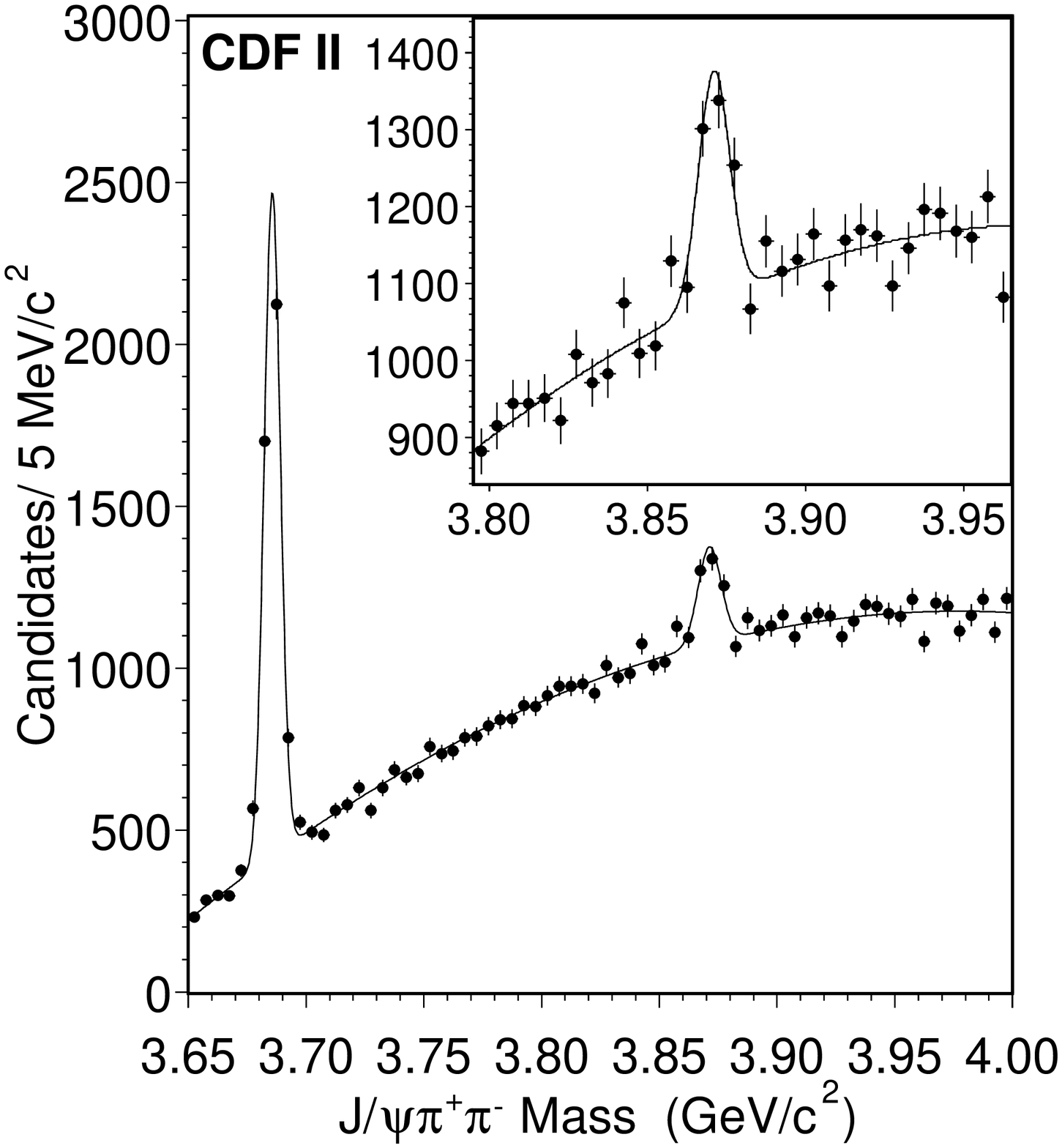}\quad 
    \includegraphics[height=3.1truein]{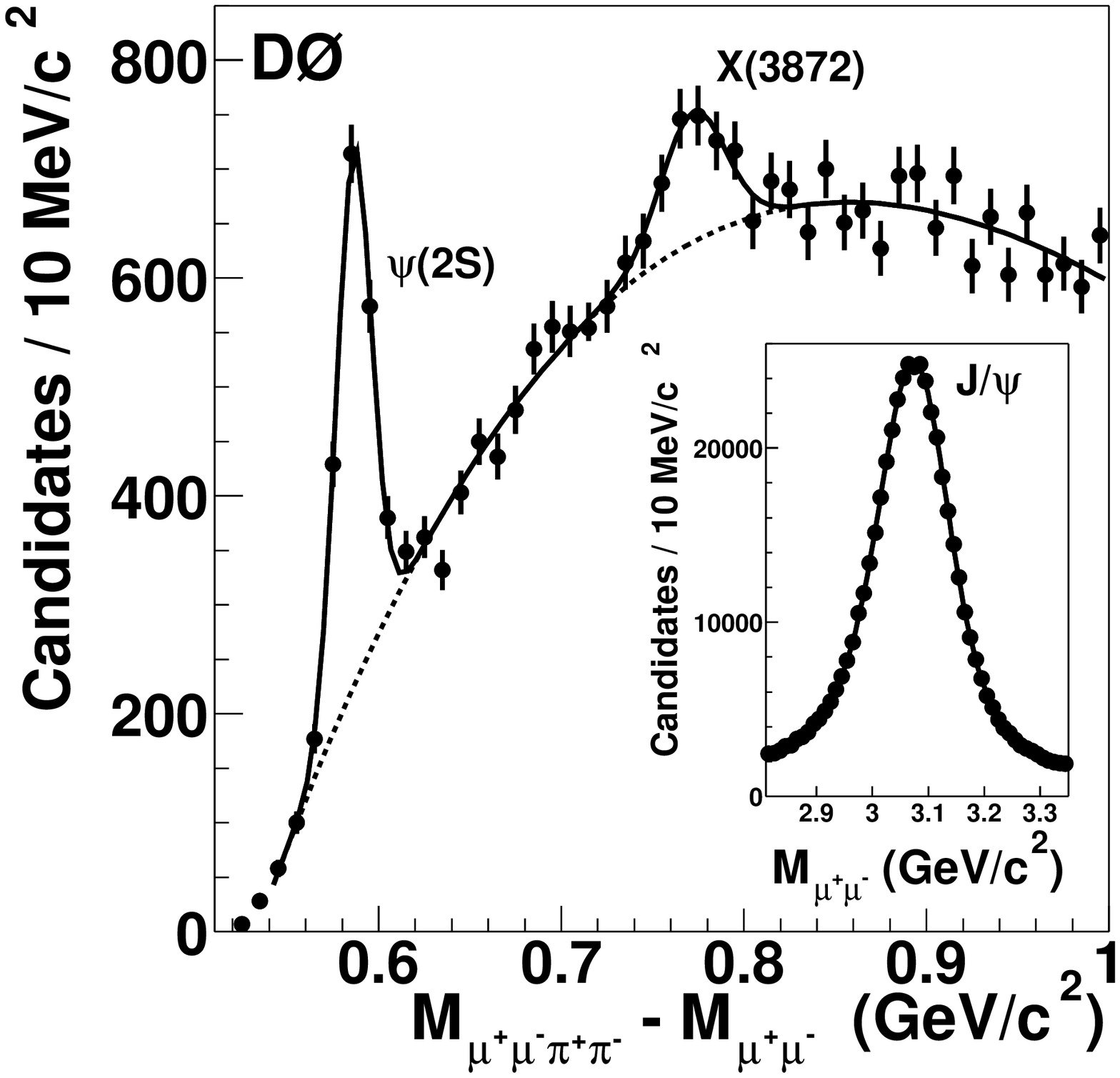}
\vspace{-30pt}
\caption{Evidence for $X(3872) \to \pi^+\pi^-\jpsi$, from
Belle~\cite{Choi:2003ue} (top left), BaBar~\cite{Aubert:2004ns} (top 
right), CDF~\cite{Acosta:2003zx} (bottom
left), and D\O~\cite{Abazov:2004kp}.  (bottom right).  The prominent peak on
the left of each panel is $\psi^\prime(3686)$; the smaller peak near
$\Delta M \equiv M(\pi^+\pi^-\ell^+\ell^-) - M(\ell^+\ell^-) \approx
775\mev, M(\jpsi\,\pi^+\pi^-) \approx 3.87\gev$ is $X(3872)$.  The CDF
and D\O\ samples are restricted to dipion masses $>500$ and $520\mev$,
respectively.}
\label{fig:X}
\end{figure*}

Last summer, Belle~\cite{Choi:2003ue} discovered $X(3872) \to 
\pi^+\pi^-\jpsi$, a candidate---by virtue of its decay mode---for a new
charmonium state. The observation was 
confirmed in short order by CDF~\cite{Acosta:2003zx}, D\O~\cite{Abazov:2004kp}, 
and  BaBar~\cite{Aubert:2004ns}. I summarize the 
observations in Figure~\ref{fig:X} and Table~\ref{table:X}.
\begin{table*}[htb]
    \caption{Observations of $X(3872) \to \pi^{+}\pi^{-}\jpsi$. The 
    Belle and BaBar data suggest that high dipion masses are 
    favored;
    CDF and D\O\ impose cuts of $M_{\pi\pi} > (500,520)\mev$, 
    respectively.}
\begin{center}
    \begin{tabular}{|cccc|}
        \hline
        Experiment & Sample & Events & Mass (MeV)  \\
        \hline
	& & &  \\[-6pt]
        Belle & 152M $\Upsilon(4\mathrm{S})\to B\bar{B}$ & $35.7 \pm 6.8$ & $3872.0 
        \pm 0.8$   \\
        CDF & $220\pb^{-1}$ & $730 \pm 90$ & $3871.4 \pm 0.8$  \\
        D\O & $230\pb^{-1}$ & $522 \pm 100$ & $3871.8 \pm 4.3$ 
	\\
        BaBar & 117M $\Upsilon(4\mathrm{S})\to B\bar{B}$ & $25.4 \pm 8.7$ & 
        $3873.4 \pm 1.4$  \\
	& & &  \\[-6pt]
	 \multicolumn{3}{|c}{Average} & {$3871.9 \pm 0.6$}  \\
         \hline
    \end{tabular}
\end{center}
\vspace*{-18pt}
\label{table:X}
\end{table*}

It is tantalizing that $X(3872)$ lies almost precisely at the
$D^0\bar{D}^{*0}$ threshold, $3871.5\mev$. Belle places an upper 
limit of $2.3\mev$ on the width of $X$. The production rates in 
2-TeV $\bar{p}p$ collisions and the similar production 
characteristics of $X$ and $\psi(2\mathrm{S})$ argue for appreciable prompt 
production at the Tevatron. A quantitative measure of prompt production
\textit{versus} $B$ decay as the source of $X$ should be forthcoming soon.

The natural prejudice is that $X(3872)$ should be identified as the 
\slj{3}{3}{2} $\psi_{2}$
charmonium state, with $J^{PC} = 2^{--}$, but this expectation 
encounters challenges: The mass is somewhat higher than 
the $3815\mev$ we expected in a single-channel potential 
model~\cite{Eichten:2002qv}, but the mismatch is diminished once we 
take account of
coupling to open-charm channels~\cite{Eichten:2004uh}.  Perhaps more serious is the fact that 
the prominent---even dominant---radiative decays, $\psi_{2} 
\to \gamma\chi_{c1,2}$ that we anticipated have not been seen. At 
90\% CL, Belle~\cite{Choi:2003ue,Choi:2004xj} limits
\begin{equation}
\mathcal{R}_{1,2} \equiv    \frac{\Gamma(X(3872) \to \gamma 
	\chi_{c1,2})}{\Gamma(X(3872) \to 
	\pi^{+}\pi^{-}\jpsi)}< 0.89, 1.1\;.
	\label{eq:radlims}
\end{equation}

The numerator is readily calculable in the framework of 
nonrelativistic quantum mechanics, but we do not have good 
theoretical control of the denominator. In the color-multipole 
expansion, the Wigner-Eckart theorem for E1-E1 transitions
predicts equal rates for all the 
$1\mathrm{D} \to 1\mathrm{S}\,\pi\pi$ cascades, but this does not take into
account kinematic differences that arise when the initial 1D states or
the final 1S states are not degenerate in mass. Moreover, the one 
rate to which we might normalize is imperfectly known.
The BES-II Collaboration reports
$\mathcal{B}(1\slj{3}{3}{1} \to \pi^+\pi^-\jpsi) = (0.338 \pm 0.137 \pm
0.082)\%$, or $\Gamma(1\slj{3}{3}{1} \to \pi^+\pi^-\jpsi) = 80 \pm 32
\pm 21\kev$~\cite{Rong:2004ei}.  This value is challenged by a CLEO-$c$
limit~\cite{Skwarnicki:2003wn}, $\mathcal{B}(1\slj{3}{3}{1} \to
\pi^+\pi^-\jpsi) < 0.26\%$ at 90\% CL. [See Ref.~\cite{Rosner:2004mi} 
for a critical assessment.]  This is a terribly hard measurement, but a
precise calibration for the 1D properties is urgently needed!

If, for illustration, we normalize to the BES-II central value, we 
expect $\mathcal{R}_{1,2} \approx 2.6,0.6$; the limit (\ref{eq:radlims}) on 
the $\gamma\chi_{c1}$ transition is a source of discomfort for the 
1\slj{3}{3}{2} interpretation.

\subsection{Alternative Assignments for $X(3872)$}
Interpretations of $X(3872)$ other than 1\slj{3}{3}{2} $c\bar{c}$ fall 
into two classes: those that attribute special significance to the 
position of $X$ at the $D^{0}\bar{D}^{*0}$ threshold, and those that 
treat the threshold as a complicating feature.

The most general threshold remark is that cusps---which may result in 
narrow resonances---are commonplace when new channels open; a 
$D\bar{D}^{*}$ $s$-wave disturbance would place $X(3872)$ as a 
$1^{++}$ state~\cite{Bugg:2004rk}.
The notion that charm molecules might be formed by attractive pion
exchange between $D$ and $\bar{D}^*$ mesons has a long history, and has
been invoked as a possible interpretation for $X(3872)$ by
T\"{o}rnqvist~\cite{Tornqvist:2004qy} and
others~\cite{Voloshin:2003nt,Wong:2003xk,Swanson:2003tb}.  A maximally
attractive channel analysis suggests that lightly bound deuteron 
analogues, should be $J^{PC} =
0^{-+}\hbox{ or }1^{++}$ states.  Symmetries forbid the decay
of these levels into $(\pi\pi)_{I=0}\jpsi$; the isospin-violating
$(\pi\pi)_{I=1}\jpsi$ mode is required. (The
$D^+$-$D^0$ and $D^{*+}$-$D^{*0}$ mass splitting means that the
molecule is not a pure isoscalar state.)  Although an isovector dipion
might account for the observed preference for high dipion masses, it
remains to be seen whether the decay rate is large enough.   T\"{o}rnqvist has suggested
that the dissociation $X(3872) \to (D^0\bar{D}^{*0})_{\mathrm{virtual}}
\to D^0 \bar{D}^0 \pi^0$ should be a prominent decay mode of a charm
molecule, with a partial width of perhaps $50\kev$.   Belle's
 limit~\cite{Abe:2003zv},
$\mathcal{B}(B^+ \to K^+ X \to K^{+} \;D^0 \bar{D}^0 \pi^0) < 6 
\times 10^{-5}$,
is perhaps an order of magnitude from challenging this expectation. 
Swanson has suggested other diagnostic decays for charm molecules~\cite{Swanson:2004pp}.

What if the $D^{0}\bar{D}^{*0}$ threshold is not the decisive element?
Hybrid states such as $c\bar{c}g$ that manifest the gluonic degrees of
freedom might also appear in the charmonium spectrum, and should be
examined as interpretations of $X(3872)$~\cite{Close:2003mb}.\footnote{For the 
production of hybrid states in $B$ decays, see 
~\cite{Chiladze:1998ti}.} It is
fair to say that dynamical calculations of hybrid-meson properties are
in a primitive state, but lattice QCD offers some guidance. Liao \& 
Manke find that the lightest hybrids should be $0^{+-}(4.7\gev)$, $1^{-+}(4.3\gev)$, 
	$2^{+-}(4.9\gev)~\cite{Liao:2002rj}$. 
The valence gluon in
the hybrid wave function leads to the speculation that the $\eta\jpsi$
mode might be quite prominent.  The Babar
experiment~\cite{Aubert:2004fc} has found no sign of $X \to \eta\jpsi$
and quoted a limit,
$\mathcal{B}(X(3872) \to \eta\jpsi) < 2 \mathcal{B}(\psi^\prime \to 
\eta\jpsi)$, that does not favor a privileged role for the $\eta\jpsi$ mode.

A less exotic possibility is that $X(3872)$ should be identified as a
charmonium level other than 1\slj{3}{3}{2}.  The 2\slj{1}{2}{1} level
has been suggested as an alternative assignment for $X(3872)$ because it
has an allowed $\pi\pi$ transition to $\jpsi$ and a hindered M1
radiative transition to the 1P levels~\cite{Pakvasa:2003ea}.
The natural-parity 1\slj{3}{3}{3} state can decay into 
$D\bar{D}$, but its $f$-wave decay is suppressed by the centrifugal 
barrier factor, so it might be narrow enough to be identified as 
$X(3872)$. We will examine 
both of these possibilities in \S\ref{subsec:oc}.

\subsection{Additional Experimental Constraints}
Where has $X$ production \textit{not} been seen?  An analysis of BES
data on the radiative return from $e^{+}e^{-}$ collisions at
$\sqrt{s}=4.03\gev$ limits $\Gamma(X \to \ell^{+}\ell^{-})\mathcal{B}(X
\to \pi^{+}\pi^{-}\jpsi) < 10\ev$ at 90\% CL~\cite{Yuan:2003yz}. A 
slightly stronger bound follow from from $15\fb^{-1}$ of 
CLEO~III data: $\Gamma(X \to \ell^{+}\ell^{-})\mathcal{B}(X \to
\pi^{+}\pi^{-}\jpsi) < 6.8\ev$ at 90\% CL, which implies $\Gamma(X \to
\ell^{+}\ell^{-}) < 0.35\kev$ for $\mathcal{B}(X \to
\pi^{+}\pi^{-}\jpsi) > 2\%$~\cite{Stoeck}. These bounds make the 
(already implausible) 3\slj{3}{1}{1} charmonium assignment unlikely, 
and does not encourage any kind of $1^{--}$ identification.

CLEO~III also has examined untagged $\gamma\gamma$ fusion, which might
be expected to excite $0^{++}, 0^{-+}, 2^{++}, 2^{-+}, \ldots$ The
absence of a signal allows them to set the limit $(2J+1)\Gamma(X \to
\gamma\gamma) \mathcal{B}(X \to \pi^{+}\pi^{-}\jpsi) <16.7\ev$ at 90\%
CL~\cite{Stoeck}.  Interpreted as charmonium, none of these states, is expected
to show a significant $\pi\pi\jpsi$ decay.  The dominant hadronic
cascades should instead be $0^{-+}  \to  \pi\pi\eta_{c}$, 
$2\slj{3}{2}{0} \to  \pi\pi1\slj{3}{2}{0}\hbox{ or }3\pi\jpsi$,
$2\slj{3}{2}{2} \to \pi\pi1\slj{3}{2}{2}\hbox{ or }3\pi\jpsi $,
$1\slj{1}{3}{2} \to \pi\pi\eta_{c}$.

Belle's discovery paper~\cite{Choi:2003ue} compares the rates of $X$
and $\psi^\prime$ production in $B$ decays,
\begin{equation}
    \frac{\mathcal{B}(B^{+}\!\!\to\!\!K^{+}X\!\!\to \!K^{+}
    \pi^{+}\pi^{-}\!\jpsi)}{\mathcal{B}(B^{+}\!\!\to\! 
    K^{+}\psi^{\prime}\!\!\to\!\!K^{+}\pi^{+}\pi^{-}\!\jpsi)} = 
    0.063 \pm 0.014.
\end{equation}

Belle~\cite{Choi:2004xj} presented the
first information about the decay angular distribution of $\jpsi$
produced in $X \to \pi^+\pi^-\jpsi$.  It does not yet determine
$J^{PC}$, but the $1^{+-}$ 2\slj{1}{2}{1} $h_c^\prime$ 
assignment is ruled out.\footnote{For more on the diagnostic
capabilities of decay angular distributions, see Jackson's  Les
Houches lectures~\cite{JDJhouches} and the recent paper on $X(3872)$ by
Pakvasa and Suzuki~\cite{Pakvasa:2003ea}.}

\section{CHARMONIUM \& OPEN CHARM}
Stimulated by Belle's discovery of $\eta_c^\prime$, my colleagues Estia
Eichten, Ken Lane, and I outlined a coherent strategy to observe
$\eta_{c}^{\prime}$ and the remaining charmonium states that do not
decay into open charm, $h_{c}(1\slj{1}{2}{1})$,
$\eta_{c2}(1\slj{1}{3}{2})$, and $\psi_{2}(1\slj{3}{3}{2})$, through
$B$-meson gateways~\cite{Eichten:2002qv}.  We argued that radiative
transitions among charmonium levels and $\pi\pi$ cascades to
lower-lying charmonia would enable the identification of these states.
Ko, Lee and Song~\cite{Ko:1997rn} discussed the observation of the
narrow 1D states by photonic and pionic transitions, and
Suzuki~\cite{Suzuki:2002sq} emphasized that the cascade decay $B
\rightarrow h_c K^{(*)} \rightarrow \gamma \eta_c K^{(*)}$ offers a
promising technique to look for $h_{c}$. The position of $X(3872)$ 
prompted us to analyze the influence of open charm on the properties of 
 charmonium levels that populate the threshold region between $2M_D$ 
 and $2M_{D^*}$~\cite{Eichten:2004uh}.

\subsection{A Coupled-Channel Model}
Our  command of quantum chromodynamics does not yet enable us to derive
 a realistic description of the interactions that communicate between
 the $c\bar{c}$ and $c\bar{q}+\bar{c}q$ sectors.  The Cornell group showed long ago that a very simple model that
 couples charmonium to charmed-meson decay channels confirms the
 adequacy of the single-channel $c\bar{c}$ analysis below threshold and
 gives a qualitative understanding of the structures observed above
 threshold~\cite{Eichten:1978tg,Eichten:1980ms}. 

 The Cornell formalism
 generalizes the $c\bar{c}$ model 
 without introducing new parameters, writing the interaction 
 Hamiltonian in second-quantized form as
 \begin{equation}
     \mathcal{H}_{I} = \cfrac{3}{8} \sum_{a=1}^{8} 
     \int\!:\!\rho_{a}(\mathbf{r}) V(\mathbf{r} - 
     \mathbf{r}^{\prime})\rho_{a}(\mathbf{r}^{\prime})\!:\! 
     d^{3}{r}\,d^{3}{r}^{\prime}\!,
     \label{eq:CCCMH}
 \end{equation}
 where $V$ is the charmonium potential and $\rho_{a}(\mathbf{r}) = 
 \cfrac{1}{2}\psi^{\dagger}(\mathbf{r})\lambda_{a}\psi(\mathbf{r})$ is the color 
 current density, with $\psi$ the quark field operator and 
 $\lambda_{a}$ the octet of SU(3) matrices. To generate the relevant 
 interactions, $\psi$ is expanded in creation and annihilation 
 operators (for charm, up, down, and strange quarks), but transitions 
 from two mesons to three mesons and all transitions that violate the 
 Zweig rule are omitted. It is a good approximation to neglect all 
 effects of the Coulomb piece of the potential in (\ref{eq:CCCMH}).

 \begin{table*}[tb]
 \caption{Charmonium spectrum, including the influence of open-charm 
 channels. All masses are in MeV. The penultimate column holds an estimate 
 of the spin splitting due to tensor and spin-orbit forces in a 
 single-channel potential model. The last 
 column gives the spin splitting induced by communication with 
 open-charm states, for an initially unsplit multiplet.
 \label{table:delM}}
 \begin{center}
 \begin{tabular}{|ccccc|}
 \hline 
 State & Mass & Centroid & $\begin{array}{c} \textrm{Splitting} 
 \\ \textrm{(Potential)} \end{array}$ &  $\begin{array}{c} \textrm{Splitting} 
 \\ \textrm{(Induced)} \end{array}$ \\
 \hline
 & & & & \\[-9pt]
 $\begin{array}{c}
     1\slj{1}{1}{0} \\
     1\slj{3}{1}{1} 
 \end{array}$ &
 $\begin{array}{c} 2\,979.9 \\ 3\,096.9 
 \end{array}$ & $3\,067.6$ & $\begin{array}{c} -90.5 
 \\ +30.2 \end{array}$
    &  $\begin{array}{c} +2.8  \\ -0.9 \end{array} $\\ & & & & \\[-6pt]
 $\begin{array}{c} 
 1\slj{3}{2}{0} \\
 1\slj{3}{2}{1} \\
 1\slj{1}{2}{1} \\
 1\slj{3}{2}{2}  \end{array}$ &  $\begin{array}{c}
 3\,415.3\\
 3\,510.5\\
 3\,525.3\\
 3\,556.2 \end{array}$
 & $3\,525.3$ &  $\begin{array}{c} 
 -114.9 \\ -11.6
 \\ +1.5 \\ -31.9 \end{array}$ & 
 $\begin{array}{c} +5.9 \\ -2.0 \\ +0.5 \\ -0.3
 \end{array}$ \\
 & & & & \\[-6pt]
 $\begin{array}{c}
     2\slj{1}{1}{0} \\
     2\slj{3}{1}{1} 
 \end{array}$ &
 $\begin{array}{c} 3\,637.7 \\ 3\,686.0
 \end{array}$ & $3\,673.9$ & $\begin{array}{c} -50.4 
 \\ +16.8 \end{array}$
    &  $\begin{array}{c} +15.7  \\ -5.2 \end{array} $\\
    & & & & \\[-6pt]
 $\begin{array}{c} 
 1\slj{3}{3}{1} \\
 1\slj{3}{3}{2} \\
 1\slj{1}{3}{2} \\
 1\slj{3}{3}{3}  \end{array}$ &  $\begin{array}{c}
 3\,769.9\\
 3\,830.6\\
 3\,838.0\\
 3\,868.3 \end{array}$
 & 
 (3\,815) & 
 $\begin{array}{c} -40 \\ 0 \\ 
 0 \\ +20 \end{array} $ & 
 $\begin{array}{c} -39.9 \\ -2.7\\ +4.2 \\ +19.0
 \end{array}$ \\ & & & & \\[-6pt]
 $\begin{array}{c} 
 2\slj{3}{2}{0} \\
 2\slj{3}{2}{1} \\
 2\slj{1}{2}{1} \\
 2\slj{3}{2}{2}  \end{array}$ &  $\begin{array}{c}
 3\,931.9\\
 4\,007.5\\
 3\,968.0\\
 3\,966.5 \end{array}$
 & 3\,968 & 
 $\begin{array}{c} -90 \\ -8 \\ 
 0 \\ +25 \end{array} $ & 
 $\begin{array}{c} +10 \\ +28.4 \\ -11.9 \\ -33.1
 \end{array}$ \\[3pt]
 \hline
 \end{tabular}
 \end{center}
 \vspace*{-18pt}
 \end{table*}
The basic coupled-channel interaction (\ref{eq:CCCMH}) is
 spin-independent, but the different energy denominators
 induce spin-dependent forces that affect the
 charmonium states.  These spin-dependent forces give rise to S-D
 mixing that contributes to the $\psi(3770)$ electronic width, for
 example, and are a source of additional spin splitting, shown in the
 rightmost column of Table~\ref{table:delM}.  To compute the induced 
 splittings, we adjust the bare centroid of the spin-triplet states so 
 that the physical centroid, after inclusion of coupled-channel 
 effects, matches the value in the middle column of 
 Table~\ref{table:delM}. 
 As expected, the shifts
 induced in the low-lying 1S and 1P levels are small.  For the other
 known states in the 2S and 1D families, coupled-channel effects are
 noticeable and interesting.  

In a simple potential picture, the $\eta_{c}^{\prime}(2\slj{1}{1}{0})$ 
 level lies below the $\psi^{\prime}(2\slj{3}{1}{1})$ by the hyperfine 
 splitting given by $M(\psi^{\prime}) - M(\eta_{c}^{\prime}) =
 32\pi\alpha_{s}|\Psi(0)|^{2}/9m_{c}^{2}$. Normalizing to the observed 
 1S hyperfine splitting, $M(\jpsi) - M(\eta_{c}) = 117\mev$, we  
 would find 
 $M(\psi^{\prime}) - M(\eta_{c}^{\prime}) = 67\mev$,
which is larger than the observed $48.3 \pm 4.4\mev$, as is typical 
for potential-model calculations.  The 2S induced
 shifts in Table~\ref{table:delM} draw $\psi^{\prime}$ and
 $\eta_{c}^{\prime}$ closer by $20.9\mev$, substantially improving the
 agreement between theory and experiment.  It is tempting to conclude 
 that the $\psi^{\prime}$-$\eta_{c}^{\prime}$ splitting reflects the 
 influence of virtual decay channels.

We peg the 1D masses to the observed mass of the 1\slj{3}{3}{1} $\psi(3770)$. In our model calculation, the coupling to open-charm
 channels increases the 1\slj{3}{3}{2}-1\slj{3}{3}{1} splitting to
 about $60\mev$, but does not fully account for the observed $102\mev$
 separation between $X(3872)$ and $\psi(3770)$.  Is it significant that
 the position of the $3^{--}$ 1\slj{3}{3}{3} level turns out to be very
 close to $3872\mev$?  For the 2P levels, we have no experimental 
 anchor, so we adjust the bare centroid so that the 2\slj{1}{2}{1} 
 level lies at the centroid of the potential-model calculation. 

The physical charmonium states are not pure potential-model
eigenstates.  To compute the E1 radiative transition rates, we must
take into account both the standard $(c\bar{c}) \to (c\bar{c})\gamma$
transitions and the transitions between (virtual) decay channels in the
initial and final states.  Our expectations for E1 decays of the
1\slj{3}{3}{2} and 1\slj{3}{3}{3} candidates for $X(3872)$ are shown in
Table~\ref{table:radtranstw}.
\begin{table*}[htb]
\caption{Calculated rates for E1 radiative decays of some 1D levels. \textit{Values in italics} result if the influence of 
open-charm channels is not included.\label{table:radtranstw}}
\begin{center}
\begin{tabular}{|cc|} 
\hline
    Transition ($\gamma$ energy in MeV)  & Partial width (keV) \\
\hline
 &  \\[-6pt]
$1\slj{3}{3}{2}(3872)\to\chi_{c2}\,\gamma(303)$ & 
 $\mathit{85} \to 45$  \\
 $1\slj{3}{3}{2}(3872)\to\chi_{c1}\,\gamma(344)$ & 
  $\mathit{362} \to 207$  \\[6pt]
    $1\slj{3}{3}{3}(3872)\to\chi_{c2}\,\gamma(304)$ & 
     $\mathit{341} \to 299$  \\[3pt]
\hline
\end{tabular}
\end{center}
\vspace*{-18pt}
\end{table*}

\subsection{Decays into Open Charm \label{subsec:oc}}

\begin{figure*}
\begin{center}
        \includegraphics[width=18pc]{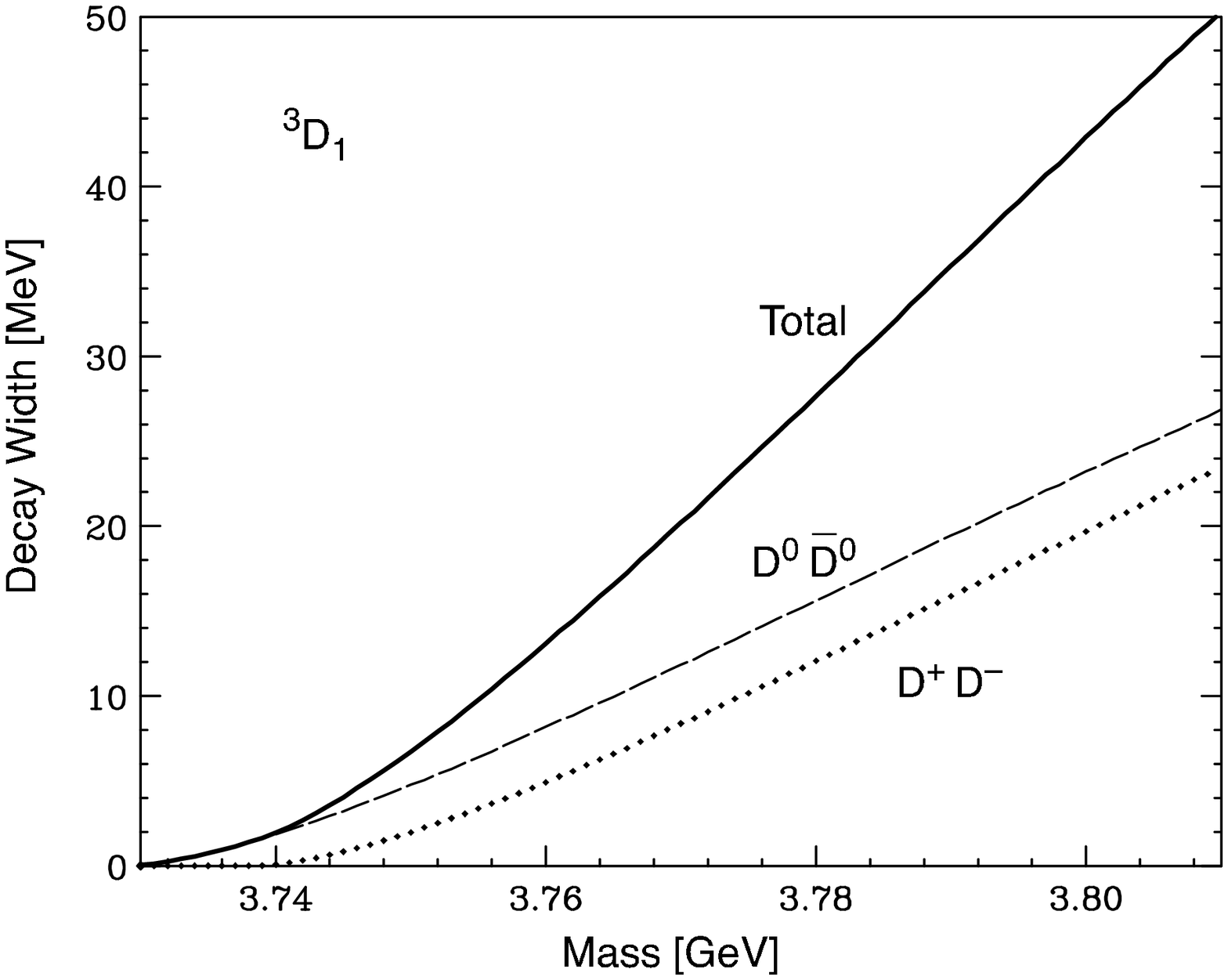}\qquad
        \includegraphics[width=18pc]{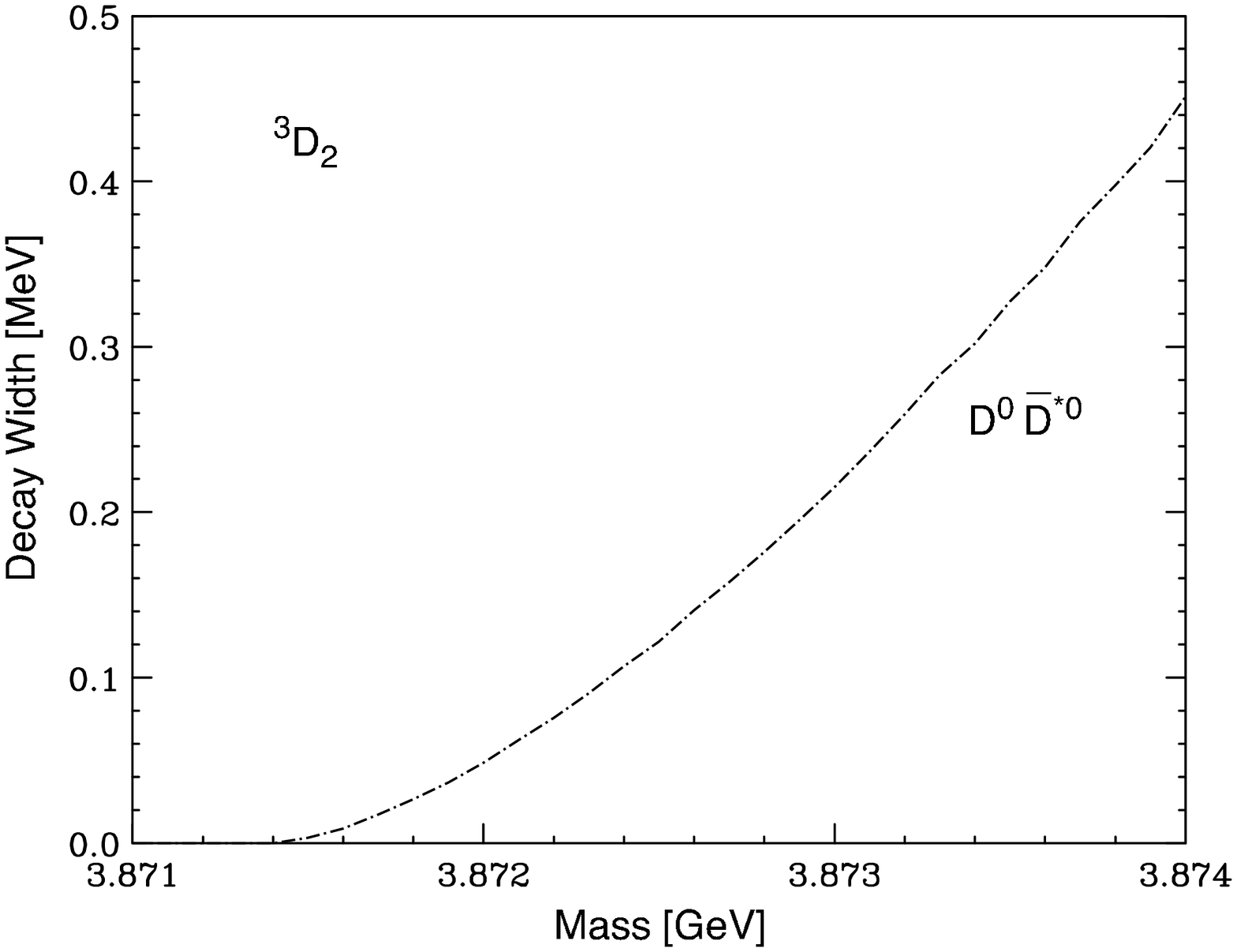}\\[3pt]
        \includegraphics[width=18pc]{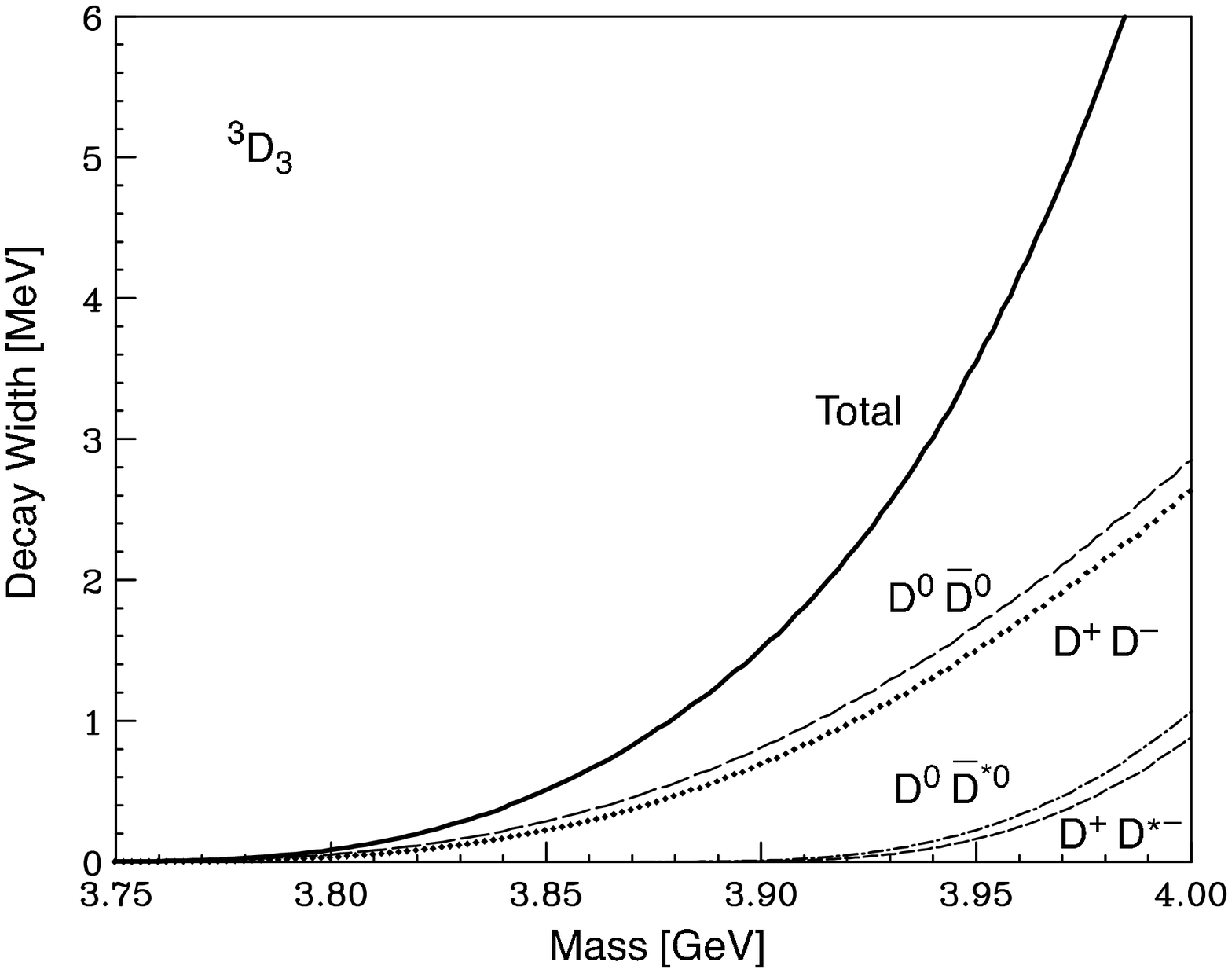}\qquad
        \includegraphics[width=18pc]{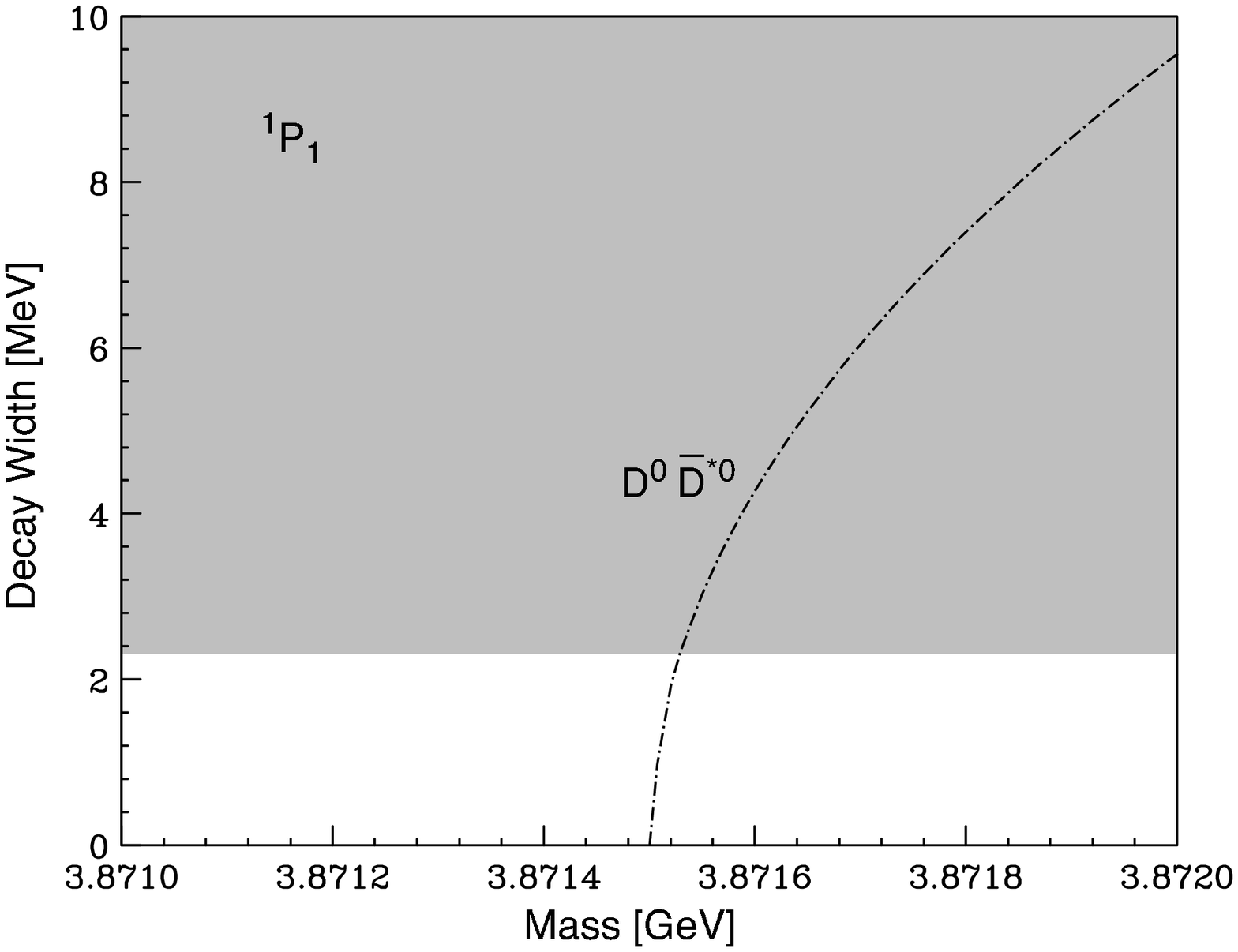}\\[3pt]
        \includegraphics[width=18pc]{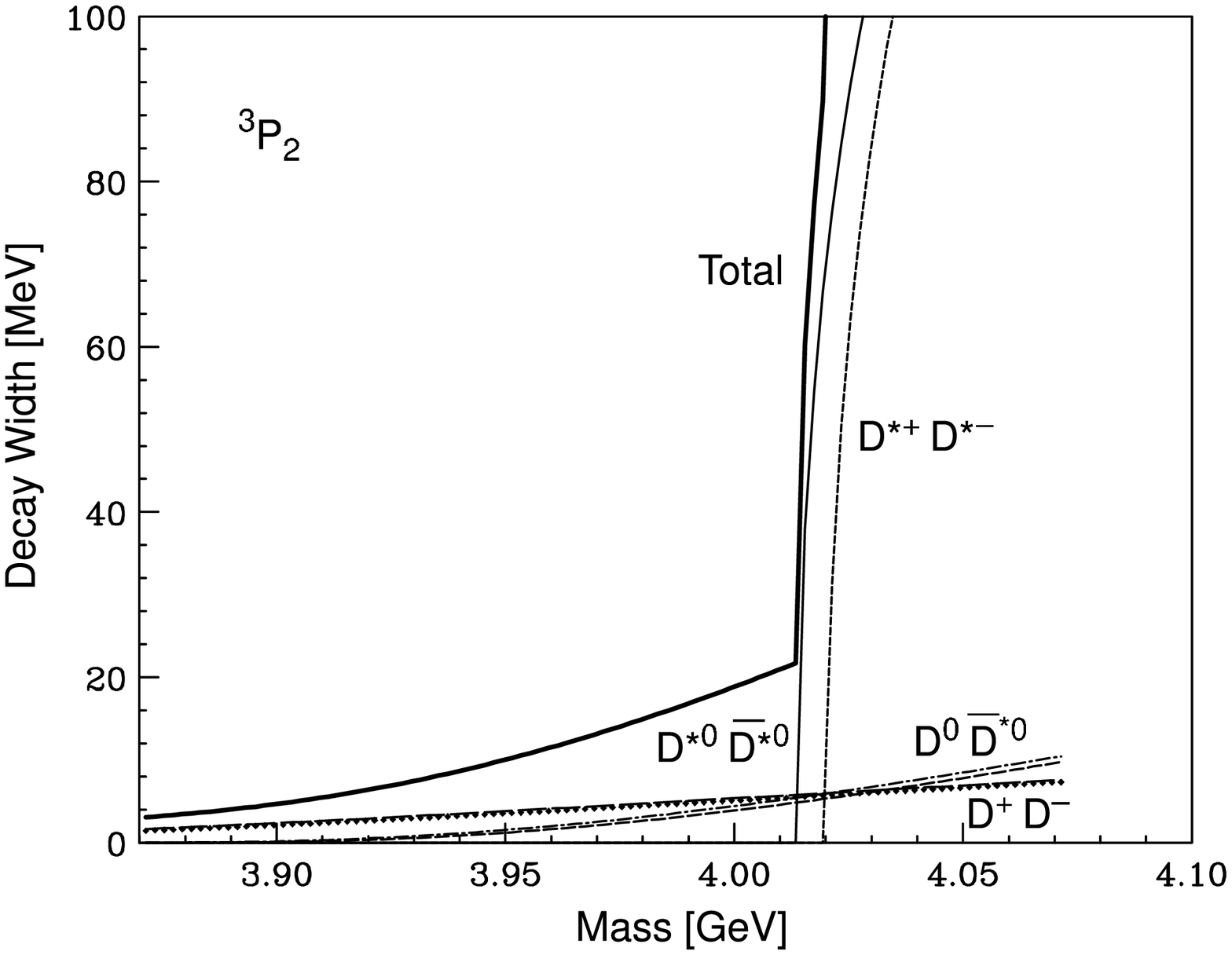}\qquad
        \includegraphics[width=18pc]{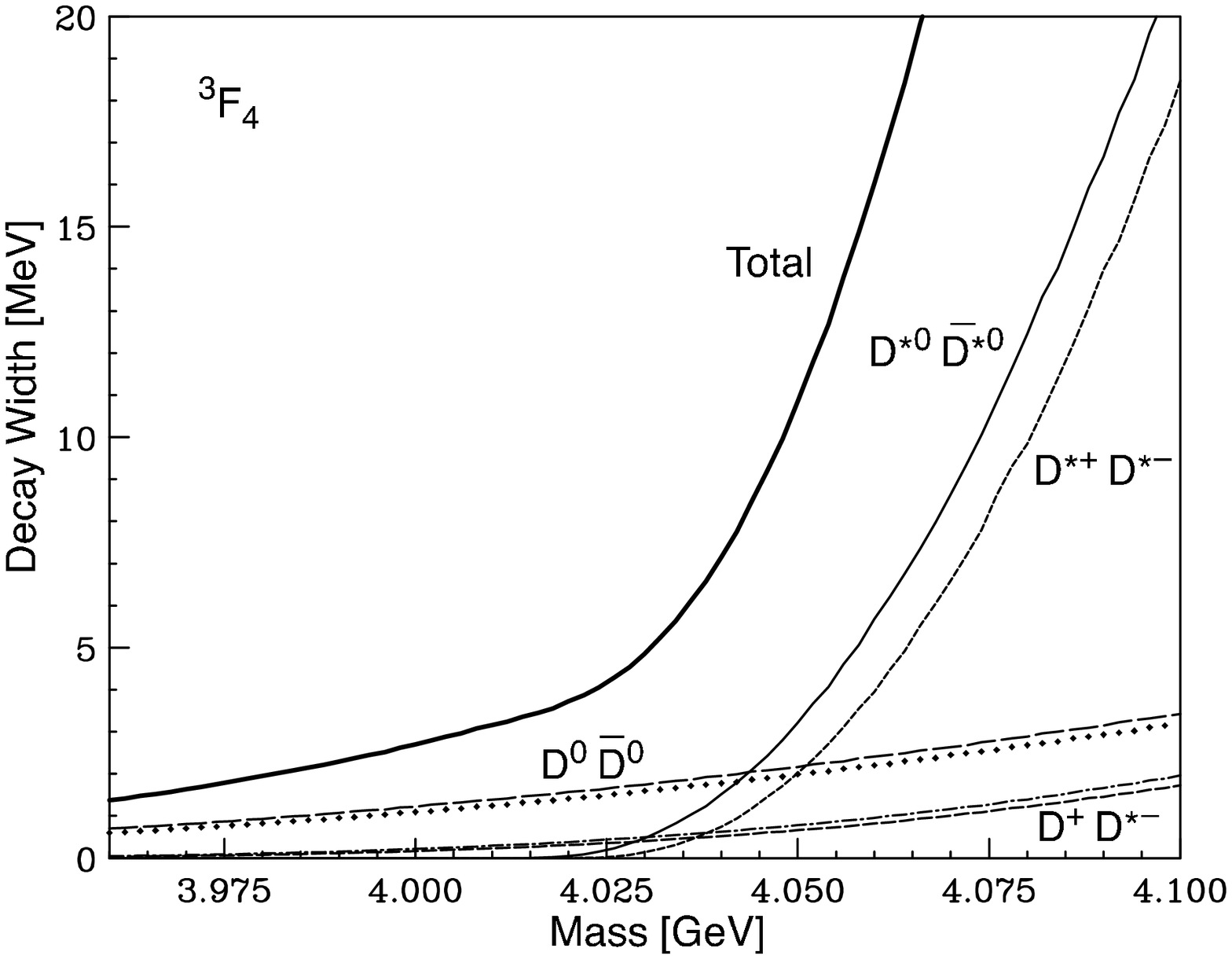}\\[3pt]
\end{center}
 \vspace*{-36pt}   
    \caption{Partial and total widths near threshold for decay of 
    charmonium states into open charm, computed in the Cornell 
    coupled-channel model. Long dashes: $D^{0}\bar{D}^{0}$, dots: 
    $D^{+}D^{-}$, dot-dashes: $D^{0}\bar{D}^{*0}$, dashes: $D^{+}D^{*-}$, 
    thin line: $D^{*0}\bar{D}^{*0}$, short dashes: $D^{*+}D^{*-}$, 
    widely spaced dots: $D_{s}\bar{D}_{s}$,
    thick line: sum of open-charm channels. Belle's 90\% C.L.\ upper 
    limit~\cite{Choi:2003ue}, $\Gamma(X(3872)) < 2.3\mev$, is 
    indicated on the \slj{1}{2}{1} window. For $D\bar{D}^{*}$ modes, 
    the sum of $D\bar{D}^{*}$ and $\bar{D}D^{*}$ is always implied.   \label{figure:OCdecays}}
\end{figure*}

Once the position of a resonance is given, the coupled-channel 
formalism yields reasonable predictions for the other resonance 
properties. The 1\slj{3}{3}{1} state $\psi^{\prime\prime}(3770)$, 
which lies some $40\mev$ above charm threshold,  
offers an important benchmark: we compute 
$\Gamma( \psi^{\prime\prime}(3770)\to 
D\bar{D}) = 20.1\mev$, to be compared with the Particle Data Group's 
fitted value of $23.6 \pm 2.7\mev$~\cite{PDBook2004}. The variation of the 
1\slj{3}{3}{1} width with mass is shown in the top left panel of 
Figure~\ref{figure:OCdecays}.\footnote{Barnes \& Godfrey~\cite{Barnes:2003vb}
estimated the decays of
several of the charmonium states into open charm, using the
\slj{3}{2}{0} model of $q\bar{q}$ production, but without carrying out
a coupled-channel analysis.}

The long-standing expectation that the 1\slj{3}{3}{2} and 1\slj{1}{3}{2} levels 
would be narrow followed from the presumption that these unnatural 
parity states should lie between the $D\bar{D}$ and $D\bar{D}^{*}$ 
thresholds, and could not decay into open charm. At $3872\mev$, both 
states can decay into $D^{0}\bar{D}^{*0}$, but the partial widths 
 are 
quite small. 
We show the variation of the 1\slj{3}{3}{2} partial width with mass in
the top right panel of Figure~\ref{figure:OCdecays}; over the region 
of interest, it does not threaten the Belle bound, 
$\Gamma(X(3872))<2.3\mev$. The range of values is quite similar to 
the range estimated for $\Gamma(1\slj{3}{3}{2} \to \pi\pi\jpsi)$, so 
we expect roughly comparable branching fractions for decays into 
$D^{0}\bar{D}^{*0}$ and $\pi^{+}\pi^{-}\jpsi$. If $X(3872)$ does turn 
out to be the 1\slj{3}{3}{2} level, we expect $M(1\slj{1}{3}{2}) = 
3880\mev$ and $\Gamma(1\slj{1}{3}{2} \to D^{0}\bar{D}^{*0}) \approx 
1.7\mev$.

The natural-parity 1\slj{3}{3}{3} state can decay into 
$D\bar{D}$, but its $f$-wave decay is suppressed by the centrifugal 
barrier factor, so the partial width is less than $1\mev$ at a mass 
of $3872\mev$. Although estimates of the hadronic cascade transitions 
are uncertain, the numbers in hand lead us to expect 
$\Gamma(1\slj{3}{3}{3} \to \pi^{+}\pi^{-}\jpsi) \ltap \cfrac{1}{4} 
\Gamma(1\slj{3}{3}{3} \to D\bar{D})$, whereas $\Gamma(1\slj{3}{3}{3} 
\to \gamma\chi_{c2}) \approx \cfrac{1}{3}\Gamma(1\slj{3}{3}{3} \to 
D\bar{D})$, if $X(3872)$ is identified as 1\slj{3}{3}{3}.
The variation of $\Gamma(1\slj{3}{3}{3} \to 
D\bar{D})$ with mass is shown in the 
middle left panel of Figure~\ref{figure:OCdecays}. Note that if 
1\slj{3}{3}{3} is not to be identified with $X(3872)$, \textit{it may still 
be discovered as a narrow $D\bar{D}$ resonance,} up to a mass of about 
$4000\mev$.

In their study of $B^{+} \to K^{+} \psi(3770)$ decays, the Belle 
Collaboration~\cite{Abe:2003zv} has set 90\% CL upper limits on the 
transition $B^{+} \to K^{+} X(3872)$, followed by $X(3872) \to 
D\bar{D}$. Their limits imply that
$    \mathcal{B}(X(3872) \!\to\! D^{0}\bar{D}^{0}) \ltap 
    4\mathcal{B}(X \!\to\! \pi^{+}\pi^{-}\jpsi)$, and
    $ \mathcal{B}(X(3872) \!\to\! D^{+}D^{-})   \ltap 
    3\mathcal{B}(X \!\to\! \pi^{+}\pi^{-}\jpsi)$.
    This constraint is already intriguingly close to the level at which we 
would expect to see $1\slj{3}{3}{3} \to D\bar{D}$.

The constraint on the total width of $X(3872)$ raises more of a 
challenge for the 2\slj{1}{2}{1} candidate, whose
$s$-wave decay to $D^{0}\bar{D}^{*0}$ rises dramatically from 
threshold, 
as shown in the middle right panel of Figure~\ref{figure:OCdecays}. 
Within the current uncertainty ($3871.7 \pm 0.6\mev$) in the mass of 
$X$, the issue cannot be settled, but the 2\slj{1}{2}{1} 
interpretation is viable only if $X$ lies below $D^{0}\bar{D}^{*0}$ 
threshold. If a light 2\slj{1}{2}{1} does turn out to be $X(3872)$, 
then its 2\slj{3}{2}{J} partners should lie nearby. In that case, 
they should be visible as relatively narrow charm-anticharm 
resonances. At $3872\mev$, we estimate  $\Gamma(2\slj{3}{2}{1}\to D\bar{D}^{*}) 
\approx 21\mev$ and $\Gamma(2\slj{3}{2}{2}\to D\bar{D}) \approx 
3\mev$. The bottom left panel in Figure~\ref{figure:OCdecays} shows 
that the 2\slj{3}{2}{2} level remains relatively narrow up to the 
opening of the $D^{*}\bar{D}^{*}$ threshold.

I point out one more candidate for a narrow resonance of charmed 
mesons: The 1\slj{3}{4}{4} level remains narrow 
($\Gamma(1\slj{3}{4}{4} \to \hbox{charm}) \ltap 
5\mev$) up to the 
$D^{*}\bar{D}^{*}$ threshold, as 
illustrated in the bottom right panel of Figure~\ref{figure:OCdecays}. Its allowed decays into $D\bar{D}$ and 
$D\bar{D}^{*}$ are inhibited by $\ell = 4$ barrier factors, whereas 
the $D^{*}\bar{D}^{*}$ channel is reached by $\ell = 2$.
%
\section{FOLLOWING UP  $X(3872)$}
The first order of experimental business is to establish the nature of
$X(3872)$.  The charmonium interpretation and its prominent rivals
require that $X(3872)$ be a neutral isoscalar.  Are there charged
partners?  In the decay $X(3872) \to \pi^{+}\pi^{-}\jpsi$, the dipion
angular distributions and the dipion mass spectrum~\cite{Ko:2004cz}
should lead to a better understanding of the $X$ quantum numbers.
Determining the $J^{PC}$ quantum numbers of $X$ is absolutely crucial
to thin the herd of candidates.  Other diagnostics of a general nature
have been discussed in
Refs.~\cite{Close:2003mb,Barnes:2003vb,Pakvasa:2003ea,Close:2003sg}.

A search for $X(3872) \to \pi^{0}\pi^{0}\jpsi$
will be highly informative.  Observing a significant $\pi^{0}\pi^{0}\jpsi$ signal
establishes that $X$ is odd under charge conjugation~\cite{Barnes:2003vb}.  
The ratio $\mathcal{R}_{0} \equiv
\Gamma(X\to \pi^{0}\pi^{0}\jpsi)/\Gamma(X \to \pi^{+}\pi^{-}\jpsi)$
measures the dipion isospin~\cite{VoloshinPC}.  Writing $\Gamma_{I}\equiv \Gamma(X \to
(\pi^{+}\pi^{-})_{I}\jpsi)$, we see that $\mathcal{R}_{0} =
\cfrac{1}{2}/(1 + \Gamma_{1}/\Gamma_{0})$, up to kinematic corrections.
Deviations from $\mathcal{R}_{0}=\cfrac{1}{2}$ signal the
isospin-violating decay of an isoscalar, or the isospin-conserving
decay of an isovector.  Radiative decay rates and the prompt (as
opposed to $B$-decay) production fraction will provide important
guidance.  

Within the charmonium framework, $X(3872)$ is most naturally 
interpreted as the 1\slj{3}{3}{2} or 1\slj{3}{3}{3} level, both of 
which have allowed decays into $\pi\pi\jpsi$. The 
$2^{--}$ 1\slj{3}{3}{2} state is forbidden by parity conservation to 
decay into $D\bar{D}$ but has a modest $D^{0}\bar{D}^{*0}$ partial width 
for masses near $3872\mev$. Although the uncertain $\pi\pi\jpsi$ 
partial width makes it difficult to estimate relative branching 
ratios, the decay $X(3872) \to \chi_{c1}\,\gamma(344)$ should show itself 
if $X$ is indeed 1\slj{3}{3}{2}. The $\chi_{c2}\,\gamma(303)$ line 
should be seen with about $\cfrac{1}{4}$ the strength of 
$\chi_{c1}\,\gamma(344)$.  In our coupled-channel calculation, the 
1\slj{3}{3}{2} mass is about $41\mev$ lower than the observed 
$3872\mev$. In contrast, the computed 1\slj{3}{3}{3} mass is quite 
close to $3872\mev$, and 1\slj{3}{3}{3} does not have an E1 
transition to $\chi_{c1}\,\gamma(344)$. The dominant decay of the 
$3^{--}$ 1\slj{3}{3}{3} state should be into $D\bar{D}$; a small 
branching fraction for the $\pi\pi\jpsi$ discovery 
mode would imply a large production rate. One radiative 
transition should be observable, with $\Gamma(X(3872) \to 
\chi_{c2}\,\gamma(303)) \gtap \Gamma(X(3872) \to \pi^{+}\pi^{-}\jpsi)$. 
I stress the importance of searching for the $\chi_{c1}\,\gamma(344)$ 
and $\chi_{c2}\,\gamma(303)$ lines.
    
It will not be easy to improve on the existing bound, 
$\Gamma(X(3872))<2.3\mev$~\cite{Choi:2003ue}, but a tighter limit or, ideally, a 
measurement, would be a very useful discriminant for theoretical 
interpretations.

Beyond pinning down the character of $X(3872)$, experiments can search
for additional narrow charmonium states in radiative and hadronic
transitions to lower-lying $c\bar{c}$ levels, as we emphasized in
Ref.~\cite{Eichten:2002qv,Eichten:2004uh}.  To underscore an obvious target: if
$X(3872)$ is 1\slj{3}{3}{3}, then $1\slj{3}{3}{2}$ lies near $3835\mev$.
Looking for additional narrow structures in the $\pi^{+}\pi^{-}\jpsi$
mass spectrum could be highly rewarding. \textit{You haven't found everything 
yet!}

The coupled-channel analysis presented in our most recent
paper~\cite{Eichten:2004uh} sets up specific targets for narrow
structures in neutral combinations of charmed mesons and anticharmed
mesons.  The most likely candidates correspond to $1\slj{3}{3}{3}$,
with $\Gamma(1\slj{3}{3}{3} \to D\bar{D}) \ltap 1\mev$;
$1\slj{3}{4}{4}$, with $\Gamma(1\slj{3}{4}{4} \to D\bar{D}) \ltap 5\mev$
for $M \ltap 2M(D^{*})$; and $2\slj{3}{2}{2}$, with
$\Gamma(2\slj{3}{2}{2} \to D\bar{D}, D\bar{D}^{*}) \ltap 20\mev$ for $M
\ltap 2M(D^{*})$.

Finally, let us not neglect the importance to charmonium spectroscopy
of establishing the $1\slj{1}{2}{1}\;h_{c}$ and confirming the quantum
numbers of the $2\slj{1}{1}{0}\;\eta_{c}^{\prime}(3638)$.

Theorists also have plenty to do.  We must improve our understanding of 
the influence of open-charm channels. Because the Cornell coupled-channel
model is only an approximation to QCD, it would be highly desirable to
compare its predictions with those of a coupled-channel analysis of the
\slj{3}{2}{0} model of quark pair production.\footnote{A preliminary
mention of work in progress by Barnes, Godfrey, \& Swanson appears in
Ref.~\cite{BarnesSJ}.} Ultimately, extending lattice QCD calculations
into the flavor-threshold region should give a firmer basis for
predictions.  The analysis we have carried out~\cite{Eichten:2004uh}
can be extended to the $b\bar{b}$ system, where it may be possible to
see discrete threshold-region states in direct hadronic production.

We need a more complete understanding of the production
of the charmonium states in $B$ decays and by direct hadronic
production.  We need to improve the theoretical understanding of
hadronic cascades among charmonium states, including the influence of
open-charm channels.
The comparison of charmonium transitions with their upsilon
counterparts should be informative.  

The outstanding theoretical challenge for the charmed molecule hypothesis 
is to understand possible production mechanisms of these apparently 
large and fragile states. The $c\bar{c}g$ hybrid-meson
hypothesis needs further development, with  specific predictions for 
the production mechanisms and properties of the states and a decision
tree to test the interpretation.

\section{OUTLOOK}
The discovery of the narrow state $X(3872) \to \pi^+\pi^-\jpsi$ gives
charmonium physics a rich and lively puzzle.  We do not yet know what
this state is.  If the most conventional interpretation as a charmonium
state---most plausibly, the 1\slj{3}{3}{2} or 1\slj{3}{3}{3} level---is
confirmed, we will learn important lessons about the influence of
open-charm states on $c\bar{c}$ levels.  Should the charmonium
interpretation not prevail, perhaps $X(3872)$ will herald an entirely
new spectroscopy.  In either event, several new charmonium states
remain to be discovered through their radiative decays or hadronic
transitions to lower $c\bar{c}$ levels.  Another set of $c\bar{c}$
states promise to be observable as narrow structures that decay into
pairs of charmed mesons.  In time, comparing what we learn from this
new exploration of the charmonium spectrum with analogous states in the
$b\bar{b}$ and $b\bar{c}$ families will be rewarding.  For all three
quarkonium families, we need to improve our understanding of hadronic
cascades.  Beyond spectroscopy, we look forward to new insights about
the production of quarkonium states in $B$ decays and hard scattering.
The rapid back-and-forth between theory and experiment is great fun, 
and I look forward to learning many new lessons!

\section{ACKNOWLEDGEMENTS}
It is a pleasure to thank the organizers of BEACH$\cdot$2004 for
assembling a rich and thought-provoking program in a superb setting.  
I thank Estia Eichten and Ken Lane for a stimulating
collaboration on the matters reported here. I am grateful to Peter 
Zweber and  Stephen Pordes for helpful comments.
\bibliography{QuiggLT04}
\end{document}